\documentclass[aps,prl,twocolumn,superscriptaddress,showpacs,preprintnumbers,amsmath,amssymb]{revtex4-2}
\usepackage{graphicx} 
\usepackage{dcolumn}  
\usepackage{xcolor}
\graphicspath{{ps}}

\usepackage{gensymb}
\usepackage{array}

\usepackage{lineno}
\usepackage{orcidlink}
\usepackage{hyperref}
\hypersetup{
    colorlinks=true,
    linkcolor=blue,
    filecolor=magenta,      
    urlcolor=blue,
    bookmarks=true,
    citecolor=blue,
}
\renewcommand{\arraystretch}{1.1}

\begin{document}

\title{\quad\\[1.0cm] Search for the decay $B_s^0\to J/\psi\pi^0$ at Belle experiment}

\noaffiliation
  \author{D.~Kumar\,\orcidlink{0000-0001-6585-7767}} 
  \author{B.~Bhuyan\,\orcidlink{0000-0001-6254-3594}} 
  \author{H.~Aihara\,\orcidlink{0000-0002-1907-5964}} 
  \author{D.~M.~Asner\,\orcidlink{0000-0002-1586-5790}} 
  \author{T.~Aushev\,\orcidlink{0000-0002-6347-7055}} 
  \author{R.~Ayad\,\orcidlink{0000-0003-3466-9290}} 
  \author{V.~Babu\,\orcidlink{0000-0003-0419-6912}} 
  \author{Sw.~Banerjee\,\orcidlink{0000-0001-8852-2409}} 
  \author{M.~Bauer\,\orcidlink{0000-0002-0953-7387}} 
  \author{P.~Behera\,\orcidlink{0000-0002-1527-2266}} 
% ...
\collaboration{The Belle Collaboration}

\begin{abstract}
We have analyzed 121.4 fb$^{-1}$ of data collected at the $\Upsilon(5S)$ resonance by the Belle experiment using the KEKB asymmetric-energy $e^+e^-$ collider to search for the decay $B_s^0\to J/\psi\pi^0$. We observe no signal and report an upper limit on the branching fraction $\mathcal{B}(B_s^0\to J/\psi\pi^0)$ of $1.21\times 10^{-5}$ at 90\% confidence level. This result is the most stringent, improving the previous bound by two orders of magnitude.
\end{abstract}

\maketitle

\tighten

{\renewcommand{\thefootnote}{\fnsymbol{footnote}}}
\setcounter{footnote}{0}
\section{I. INTRODUCTION}
\begin{figure}[!h]
    \centering
    \includegraphics[width=0.60\linewidth]{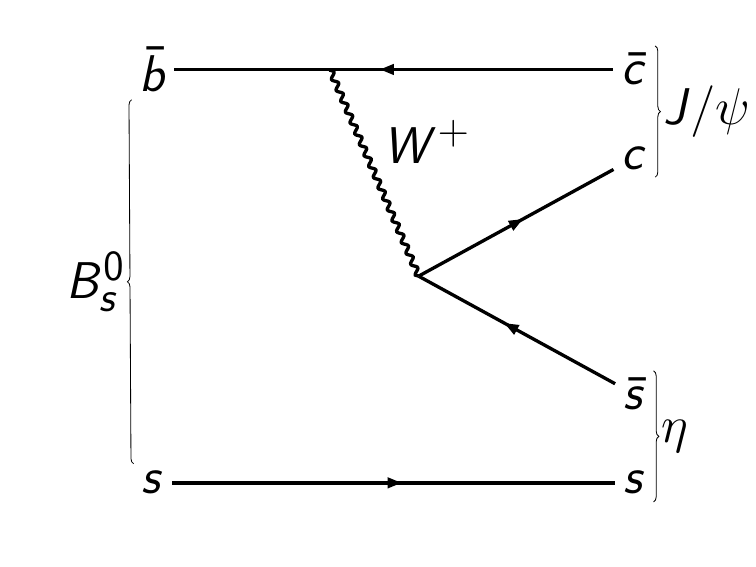}
    \caption{Tree-level Feynman diagram for the $B_s^0\rightarrow J/\psi \eta$ transition.}
    \label{fig:feynman}
\end{figure}

Understanding the physics governing the rare decay of heavy-flavored $B$-mesons is crucial to test the standard model (SM) at low energy scales. Any experimental deviation from the SM expectation would hint at new physics (NP) effects beyond the SM. At leading order, the branching fraction of the decay $B_s^0\to J/\psi\pi^0$, $\mathcal{B}(B_s^0\to J/\psi\pi^0)$, can be predicted from the measurement of the branching fraction of $B_s^0\to J/\psi\eta$ \cite{pdglive}, with a suppression factor of order $\mathcal{O}(10^{-2})$ due to the violation of strong-isospin in the $\eta-\pi^0$ transition \cite{voloshin,qhe}. Figure \ref{fig:feynman} shows the Feynman diagram for the $B_s^0 \to J/\psi \eta$ transition at the tree level, where the $\eta$ can transit to $\pi^0$ under the assumption of isospin-zero admixture in $\pi^0$. The isospin suppression factor is naively predicted from the measured ratio of the decay widths for $\psi'\to J/\psi \pi^0$ and $\psi'\to J/\psi \eta$ transitions \cite{ablikim} and corresponding theoretical prediction for $\Upsilon(2S)\to \Upsilon(1S) \pi^0$ and $\Upsilon(2S)\to \Upsilon(1S) \eta$ transitions. We expect the $\mathcal{B}(B_s\to J/\psi\pi^0)$ of the order of $4\times 10^{-6}$ due to $\eta-\pi^0$ transition. The contributions from $W$ exchange and annihilation processes are much smaller, of the order of $10^{-8}$ or less, since the gluonic production of a $\pi^0$ violates the isospin. The predictions are based on the measured branching
fractions of $B^0_d\to K^+ K^-$ and $B_s^0\to \pi^+\pi^-$ \cite{weakannihilationsmall}, which at leading order can only proceed via such transitions. 

The existing experimental limit on $\mathcal{B}(B_s^0\to J/\psi\pi^0)$ of $1.2 \times 10^{-3}$ at 90\% confidence level was first set by the L3 collaboration in 1997 \cite{l3report}. In this article, we report the search results for the $B_s^0\to J/\psi\pi^0$ decay based on $121.4$ fb$^{-1}$ of data collected by the Belle detector. Unless otherwise stated, the inclusion of the charge-conjugate decay mode is implied throughout this paper.
\section{II. The Belle Detector and The Data Sample}
The Belle detector is a cylindrical large-solid-angle magnetic spectrometer situated at the interaction point (IP) of the KEKB $e^+e^-$ beam collider \cite{Kurokawa:2001nw}. The detector consists of an innermost silicon vertex detector (SVD) followed by 50 layers of multi-wired central drift chamber (CDC) measuring the vertices, momentum, and energy loss $(dE/dx)$ of charged particles; an array of aerogel threshold Cherenkov counters (ACC) and barrel-like arrangement of time-of-flight (TOF) scintillation counters dedicated to the charged particle\textquotesingle s identification; an electromagnetic calorimeter (ECL) for the measurement of energy deposited by the charged particles and photons through electromagnetic interactions; a superconducting solenoid housing the sub-detectors in a uniform magnetic field of strength 1.5 T; and an outermost $K_L^0$ and muon (KLM) detector to identify the relatively long-lived $K_L^0$ mesons and muons. The $z$-axis of the detector points in the direction opposite to the positron beam. A detailed description of the Belle detector is given elsewhere \cite{Belle:2000cnh}.

The total $b\bar{b}$ production cross-section in $e^+e^-$ collisions at the center-of-mass (c.m.) energy of 10.86 GeV is measured to be $\sigma_{b\bar{b}}^{\Upsilon(5S)} = 0.340 \pm 0.016$ nb \cite{Belle:bbproductionfract}. A fraction $f_s = \left( 22.0 ^{+2.0} _{-2.1}\right)\%$ \cite{fsvalue} of $b\bar{b}$ events produce the kinematically allowed $B_s^{0^{(*)}}\bar{B}_s^{0^{(*)}}$ pairs, such as $B_s^{0^{*}}\bar{B}_s^{0^{*}}$, $B_s^{0^{*}}\bar{B}_s^0$, and $B_s^0\bar{B}_s^0$, with a relative percentage of $f_{B_s^{0^{*}}\bar{B}_s^{0^{*}}} = (87.0\pm 1.7)\%$ and $f_{B_s^{0^{*}}\bar{B}_s^0} = (7.3\pm 1.4)\%$ \cite{Belle:bbproductionfract}. Excited $B_s^0$ mesons decay to the ground state by emitting a low-energy photon, often undetected due to poor reconstruction efficiency. The number of $B_s^0\bar{B}_s^0$ pairs analyzed is estimated to be $N_{B_s^0\bar{B}_s^0} = 121.4\text{ fb}^{-1}\times\sigma_{b\bar{b}}^{\Upsilon(5S)}\times f_s$ = $(9.08 ^{+0.94} _{-0.98})\times 10^6$. 
\section{III. EVENT SELECTION}
We perform a blind analysis using Monte Carlo (MC) simulated events to optimize the $B_s^0\to J/\psi\pi^0$ selection criteria. The MC events are generated using \texttt{EvtGen} \cite{evtgen} followed by the \texttt{GEANT3} \cite{geant3} simulation to model the detector response. The final-state radiations are incorporated using the \texttt{PHOTOS} package \cite{photos}.

The event selection proceeds by reconstructing $J/\psi$ using the $e^+ e^-$ and $\mu^+\mu^-$ decay channels and the $\pi^0$ from two-photon final states. $J/\psi$ candidates are formed by combining two oppositely charged particles whose closest approach to the nominal IP is within 0.5 cm and 3 cm along the radial and z-axis, respectively. Electrons are identified based on the position matching between the extrapolated charged track and ECL cluster, the $dE/dx$ measurements of a charged particle in CDC, the ratio of deposited energy in ECL to the measured momentum using the tracking detector, the transverse spread of the electromagnetic shower in ECL, and light yield in ACC. In addition, the four momenta of photons within 50 mrad of the momentum direction of the selected electron track at IP are added to the electron candidate to correct the possible energy loss due to bremsstrahlung radiation.  Muons are selected using the information on penetration depth and lateral spread of the charged-particle hits in the KLM. The reconstructed invariant masses $M_{\mu^+\mu^-}$ and $M_{e^+e^-}$ are required to satisfy $ -72\text{ MeV}/c^2 < M_{\mu^+\mu^-} - m_{J/\psi} < +41 \text{ MeV}/c^2$ and $ -102\text{ MeV/}c^2 < M_{e^+e^-} - m_{J/\psi} < +47 \text{ MeV/}c^2$, respectively, where the intervals correspond to approximately $3\sigma$ region around the nominal $J/\psi$ mass, $m_{J/\psi}$ \cite{pdglive}. Asymmetric mass windows account for the radiative effects resulting in a tail towards the lower values of invariant masses. A simultaneous mass-vertex constrained fit is further imposed on the selected $J/\psi$ candidates to improve the momentum resolution. The $\chi^2$ of the fit is required to be less than 60.

The ECL clusters not matched to any tracks in CDC are identified as photons for $\pi^0\to\gamma\gamma$ reconstruction. Photon candidates must have energies greater than the threshold of 50 and 100 MeV in the barrel and end-cap regions, respectively. The barrel region covers the laboratory polar angle of $32\degree < \theta < 130\degree$, whereas end-cap regions cover the ranges $12\degree < \theta < 32\degree$ and $130\degree < \theta < 157\degree$. The $\gamma\gamma$ invariant mass, $M_{\gamma\gamma}$, is required to be within $80-180 \text{ MeV}/c^2$. We further perform a mass-constrained fit to improve the momentum resolution, and the $\pi^0$ candidates with $\chi^2<30$ on the fit quality are selected for the $B_s^0\to J/\psi\pi^0$ reconstruction.

In an event, four momenta of the selected $J/\psi$ and $\pi^0$ candidates are added to reconstruct the $B_s^0$ candidates. We compute the kinematical observables: beam-constrained mass, $M_\text{bc} = \left(\sqrt{{E_\text{beam}}^2 - {|\Vec{p}_{B_s^0}}|^2c^2}\right)/c^2$, and the energy difference, $\Delta E = E_{B_s^0} - E_\text{beam}$; where $E_\text{beam}$ is the beam energy, $\Vec{p}_{B_s^0}$ and $E_{B_s^0}$ are respectively the momentum and energy of the $B_s^0$ candidates, all calculated in the c.m. frame. The $\Delta E$ distributions for the signal candidates in three different $B_s^{0^{(*)}}\bar{B}_s^{0^{(*)}}$ production modes have slightly different means, resulting in a degradation of $\Delta E$ resolution. As a result, we define a new variable, $\Delta E^\prime = \Delta E + (M_\text{bc} - m_{B_s^0})c^2$, where $m_{B_s^0}$ is the nominal $B_s^0$ mass \cite{pdglive}. $B_s^0$ candidates with $M_\text{bc}>5.35$ GeV/$c^2$ and $ -0.2 \text{ GeV } < \Delta E'<0.1$ GeV are retained for further analysis. 

Backgrounds arising from the ``continuum'' ($e^+e^-\to q\bar{q}$, $q = u,\text{ }d,\text{ }s,\text{ and }c$); $B_s^{0^{(*)}}\bar{B}_s^{0^{(*)}}$ (referred as $bsbs$); and $B^{(*)}\bar{B}^{(*)}$,  $B^{(*)}\bar{B}^{(*)}\pi$, and $B\bar{B}\pi\pi$ (referred as non-$bsbs$) decay at $\Upsilon(5S)$ resonance are studied using a dedicated MC sample that is six times larger than the data sample. In contrast to the $B_s^0\bar{B}_s^0$ pairs produced with relatively small momenta in the c.m. frame,  the particles from the continuum background have back-to-back jet-like distributions.  This topological difference is used to suppress the continuum background by requiring the ratio of second to zeroth Fox-Wolfram moments, calculated using all the charged particles in an event, to be less than 0.4 \cite{fwmoments}.

Having applied all the aforementioned criteria, approximately $2.3$\% of the events have multiple $B_s^0$ candidates. In such cases, the candidate with the least $\chi^2$ sum of the $J/\psi$ mass-vertex-constrained fit and $\pi^0$ mass constraint fit is kept as the best candidate. Based on the MC study, we find that this procedure selects the correct $B_s^0$ candidates about 79\% and 77\% of the time in the electron and muon channels, respectively. The fraction of incorrectly reconstructed candidates in the signal events, where at least one of the $B_s^0$ daughters is either mis-reconstructed or originates from the other $B_s^0$ that accompanies the signal decay, is negligibly small ($<2\%$); hence, such events are not treated separately. From the MC simulation, the signal reconstruction efficiency is estimated to be (31.0 $\pm$ 0.1)\%, where the uncertainty is statistical only.
\section{IV. MAXIMUM LIKELIHOOD FIT}
We extract the signal yield using a two-dimensional (2D) unbinned extended maximum likelihood fit to the $M_\text{bc}$ and $\Delta E'$ distributions. The likelihood function is defined as,
\begin{equation}
    \mathcal{L}_{\rm fit} = e^{-\sum\limits_{j} n_j} \prod_{i}^{N} \left( \sum_{j} n_j P_j(M_{\text{bc}}^i, \Delta E'^i) \right),
\end{equation}
where $P_j$ are the signal and background probability distribution functions (PDFs) with a corresponding yield of $n_j$, and $N$ is the total number of data points. 
\subsection{A. The signal and background fit functions}
The PDF for the signal $M_\text{bc}$ distribution determined from a large MC simulation has three peaks attributed to $B_s^{0^{(*)}}\bar{B}_s^{0^{(*)}}$ states: the component from $B_s^{0^{*}}\bar{B}_s^{0^{*}}$ is parameterized using a Crystal Ball (CB) \cite{CBfunction} function, while the distributions from $B_s^0\bar{B}_s^{0^{*}}$ and $B_s^0\bar{B}_s^0$ are parameterized using two Gaussian (G) functions with a common mean. The final PDF for the signal $M_\text{bc}$ distribution is determined by adding the CB component with the two Gaussian components using the measured fractions of $f_{B_s^{0^*}\bar{B}_s^{0^*}}$ and $f_{B_s^{0^*}\bar{B}_s^{0}}$ \cite{Belle:bbproductionfract}. $\Delta E'$ distribution for the signal is fitted with a combination of CB and G, sharing a common mean. The linear correlation coefficient between $M_\text{bc}$ and $\Delta E'$ is small (3\%) around the prominent peaks. Therefore, a 2D PDF is calculated by multiplying the PDFs for $M_\text{bc}$ and $\Delta E'$ distributions. Backgrounds from $bsbs$ and non-$bsbs$ decays with a correctly reconstructed $J/\psi$ candidate are categorized into three types: $B_s^0\to c\bar{c} X$, $B_d^0 \to J/\psi\pi^0$, and $B\to c\bar{c} X$. For the first two types, we construct the non-parametric 2D histogram PDFs for $M_\text{bc}$ and $\Delta E'$ distributions using the MC event samples. Backgrounds from the continuum production and $B\to c\bar{c} X$ are fitted with an ARGUS \cite{argusf} function having an endpoint at 5.433 GeV/$c^2$ for the $M_\text{bc}$ distribution and first-order Chebychev polynomial for the $\Delta E'$ distribution. The final PDF is constructed by adding the signal and different background components. Based on the MC simulation, the yields corresponding to the $B_s^0\to c\bar{c} X$ and $B_d^0 \to J/\psi\pi^0$ background components are fixed to $4.33 \pm 2.08 ~ ^{+0.51}_{-0.52}$ and $5.17 \pm 2.27 ~ ^{+0.49}_{-0.53}$ events, respectively, where the first and second uncertainties are statistical and systematic uncertainties. The uncertainty on $f_s$ dominates the systematic uncertainty for $B_s^0\to c\bar{c} X$ \cite{fsvalue}, whereas the uncertainty on the branching fraction and the production fraction of $B\bar{B}X$ events at $\Upsilon(5S)$ resonance dominates the systematic uncertainty for $B_d^0\to J/\psi \pi^0$ \cite{pdglive}. All the signal PDF parameters and the background ARGUS function endpoint are fixed to the best-fit values obtained from the MC simulated events, whereas the other background parameters and the yields for the signal and remaining background component, a total of four parameters, are floated.
\subsection{B. MC simulation validation}
\begin{figure}[!h]
    \centering
    \includegraphics[width=0.9\linewidth]{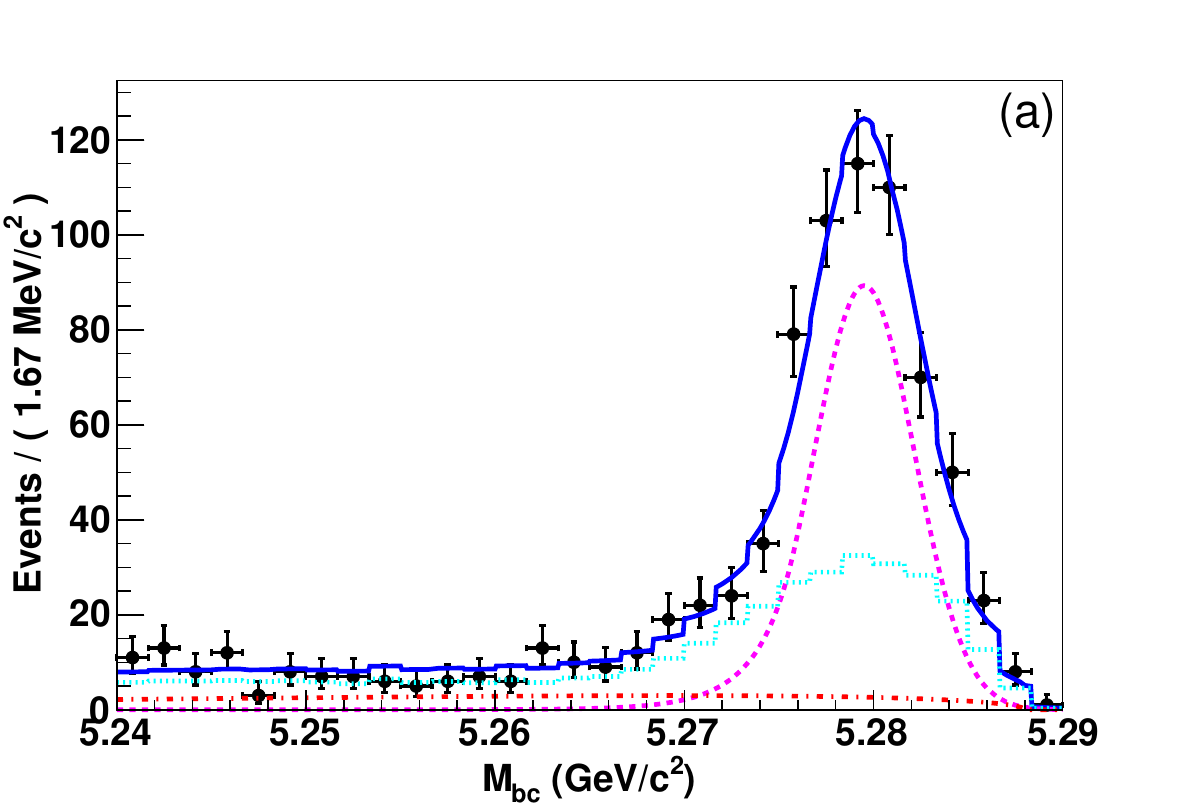}
    \includegraphics[width=0.9\linewidth]{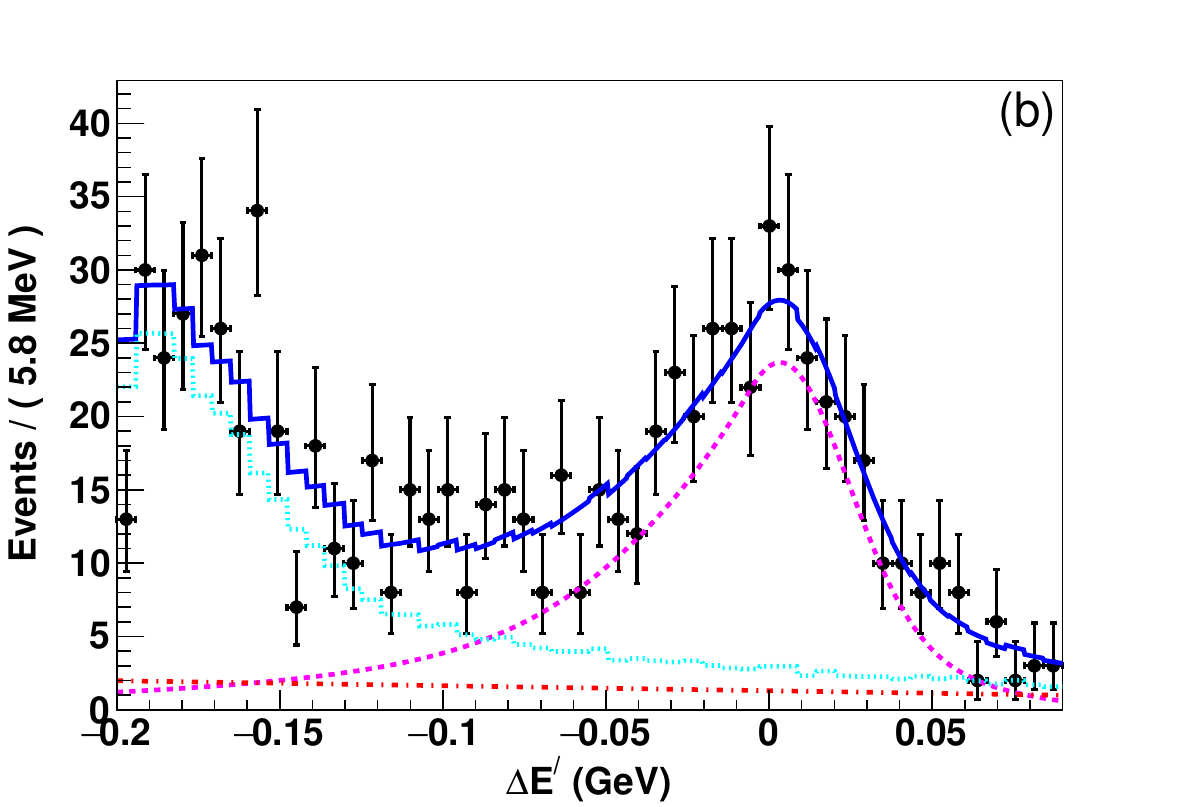}
    \caption{Projections of the fit to selected $B_d^0\to J/\psi\pi^0$ events in $e^+e^-$ collision data having (619 $\pm$ 9)$\times 10^6$ $B\bar{B}$ mesons at the $\Upsilon(4S)$ resonance: (a) $M_\text{bc}$ and (b) $\Delta E'$ distributions. Black points with error bars represent the data. Dashed (magenta), dotted (cyan), dash-dotted (red), and solid (blue) lines are the signal, $b\to c\bar{c} q$, continuum, and total PDFs, respectively.}
   \label{fig:csdata}
\end{figure}

We study the $B_d^0 \to J/\psi\pi^0$ decay at the $\Upsilon(4S)$ resonance as a control sample to validate the event selection criteria and estimate the discrepancy between the simulated and recorded data. After applying an identical set of event selection criteria except for the requirement of $M_\text{bc}>5.24$, we perform a 2D fit to the $M_\text{bc}$ and $\Delta E'$ distributions of the selected $B_d^0$ candidates, where the $\Delta E'$ for control sample is defined using the nominal mass of the $B_d^0$ meson \cite{pdglive}. The signal PDF consists of a CB function for the $M_\text{bc}$ distribution and a combination of CB and G with a common mean for the $\Delta E'$ distribution. The background from $B$ decays is dominated by the $B\to c\bar{c}X$ events. The $M_\text{bc}$ and $\Delta E'$ distributions for such events are modeled using a non-parameterized 2D histogram PDF determined from an MC simulated $B\to c\bar{c}X$ event sample, which contains 100 times more events than expected events in data. The remaining background is from the continuum production. The $M_\text{bc}$ distribution from the continuum background is parameterized using an ARGUS function with an endpoint at 5.289 GeV$/c^2$ and the $\Delta E'$ distribution is fitted with a first-order Chebychev polynomial. The means and resolutions of the signal PDF and the signal and background yields are floated in the final fit. The power-law tail-end parameters of the CB functions are fixed to the estimated values using the MC simulated signal data. The results of the unbinned extended maximum likelihood fit to the data sample are shown in Fig.~\ref{fig:csdata}. We observe $369.0 \pm 26.6$ signal events with an efficiency of $(30.9 \pm 0.1)\%$ and measure the branching fraction of $B_d^0 \to J/\psi\pi^0$ to be $(1.63 \pm 0.13) \times 10^{-5}$ (where the uncertainty is statistical only), which is in an excellent agreement with our previous result \cite{bpal:bdtojpi0}. The control sample study demonstrates excellent data and MC agreement in the signal PDF parameters which are within the statistical uncertainties. The small difference is considered as a source of additional systematic uncertainty.
\subsection{C. Fit validation}
We estimate the fit bias in the signal yield using ensembles of 5000 pseudo-samples each, where $B_s^0 \to J/\psi \pi^0$ signals are selected randomly from the MC simulated events, and the expected background events are generated using the background PDFs based on the MC simulation. The statistical fluctuations in the number of signal and background events are incorporated using the Poisson distribution. We then perform the 2D unbinned extended maximum likelihood fit to each pseudo-sample, extracting the signal yield and computing the deviation from the input signal events before incorporating the statistical fluctuations. We do not observe any significant biases for signal yield around the upper limit value. As a result, we take the mean (+5.5\%) of the pull (the difference between the fitted signal yield and the input value divided by the statistical uncertainty) distribution with input signals equal to the upper limit value as the corresponding systematic uncertainty.
\subsection{D. Fit results}
\begin{figure}[!h]
    \centering
    \includegraphics[width=0.9\linewidth]{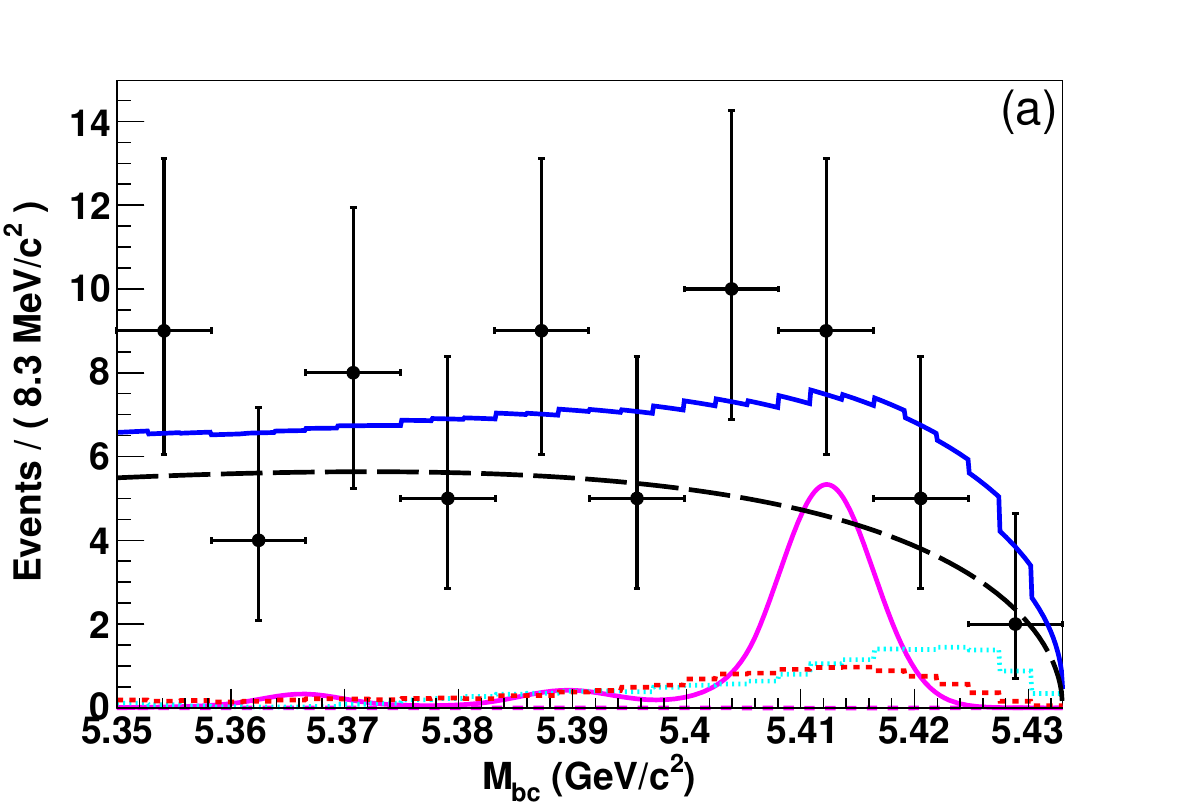}
    \includegraphics[width=0.9\linewidth]{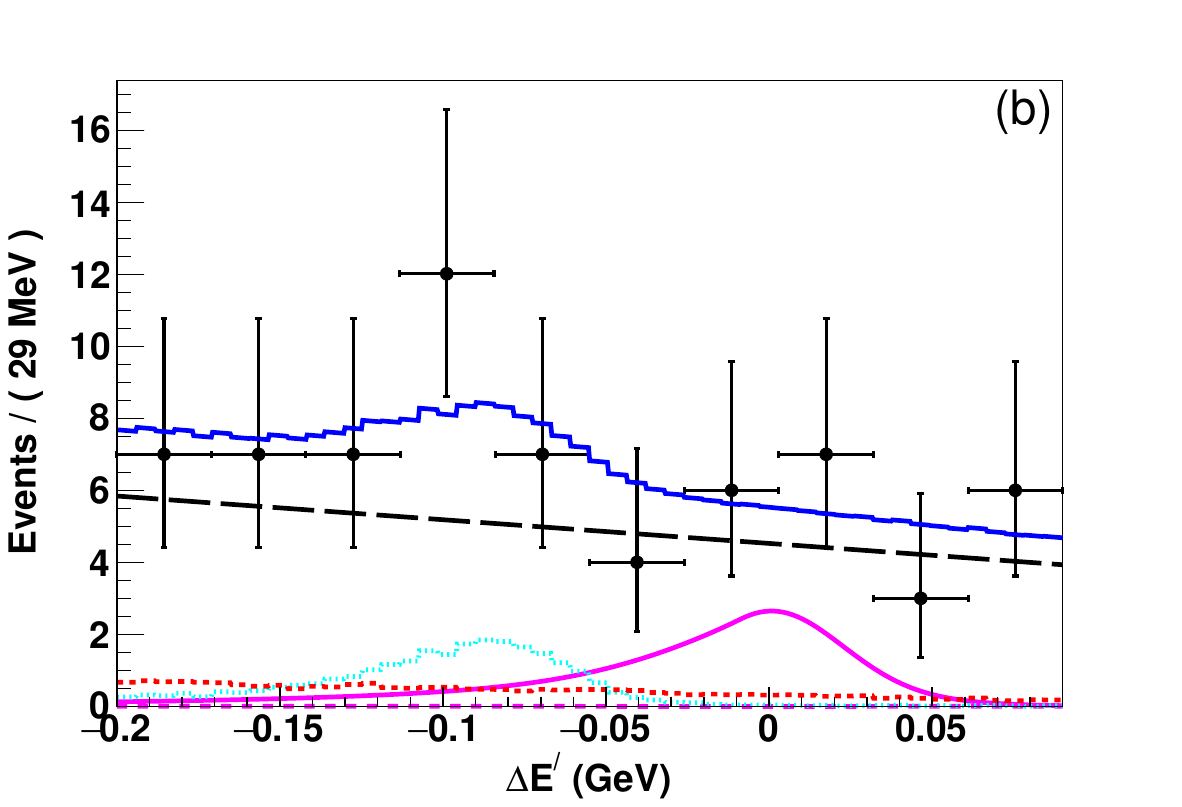}
    \caption{Projections of the fit to selected $B_s^0\to J/\psi\pi^0$ events in the 121.4 fb$^{-1}$ of $e^+e^-$ collision data at the $\Upsilon(5S)$ resonance: (a) $M_\text{bc}$ and (b) $\Delta E'$ distributions. Black points with error bars represent the data. Long-dashed (magenta), dashed (red), dotted (cyan), broken-line (black), and solid (blue) lines are the signal, $B_s^0\to c\bar{c}X$, $B_d^0\to J/\psi \pi^0$, a continuum with remaining $B\to c\bar{c} X$, and total PDFs, respectively. The signal contribution is approximately zero, whereas the distribution corresponding to 90\% CL is shown using the solid (pink) line.}
   \label{fig:realdata}
\end{figure}

Figure~\ref{fig:realdata} shows the projections of the 2D fit to the selected $B_s^0\to J/\psi\pi^0$ events in the 121.4 fb$^{-1}$ of $e^+e^-$ collision data. We obtain the yields of $0.0 \pm 3.2$ signal events and $50.0 \pm 4.0$ continuum and $B\to c\bar{c}X$ events, where the uncertainties are statistical only.

\section{V. SYSTEMATIC UNCERTAINTIES}
Except for the means and resolutions of the signal PDF, the parameters fixed to the best-fit values are varied within $\pm1\sigma$ of the statistical uncertainties to estimate the associated systematic uncertainty due to pre-determined values. The systematic uncertainty due to the slight discrepancy in the simulated and actual data is evaluated by changing the means and widths according to their correction factors from the control sample. Systematic uncertainties from the fixed widths are evaluated by smearing them with their respective ratio of widths from actual and MC data distributions in the control sample. The systematic uncertainties due to the fixed yields for $B_d^0\to J/\psi \pi^0$ and $B_s^0\to c\bar{c}X$ components are computed by varying the corresponding estimates by the total uncertainties on the fixed yields. The deviations of the signal yield from the central value of $0.0$ events are determined independently for all the fixed parameters. These deviations are then added in quadrature, assigning a systematic uncertainty of $^{+0.7}_{-0.8}$ events associated with PDF parameterization, as shown in Table \ref{table:additive}. We assign the systematic uncertainty of +0.2 events from the fit bias of +5.5\% in the signal pull. These uncertainties affect the significance of the observed signal peak and branching fraction measurements. Therefore, they are treated as additive in nature.  In contrast, the uncertainties mentioned in Table \ref{table:multiplicative} affect the branching fraction measurement through the uncertainties in signal reconstruction efficiency, the number of $B_s^0$ mesons, and branching fractions $\mathcal{B}(J/\psi\to l^+l^-)$  and $\mathcal{B}(\pi^0\to \gamma\gamma)$. These uncertainties are treated as multiplicative systematic uncertainties. The uncertainty due to the $\pi^0$ reconstruction is obtained from the study of $\tau^-\rightarrow \pi^0\pi^-\nu_\tau$ decays \cite{pi0systematic}. We assign the corresponding systematic uncertainty of 2.2\%, which is dominated by the uncertainty on the number of $\pi^0$ events, the uncertainty of the $\pi^\pm$ identification used to reconstruct the $\tau^- \to \pi^0\pi^-\nu_\tau$ events, and the uncertainty of the lepton identification used to reconstruct the decay $\tau^- \to l^- \bar{\nu}_l \nu_\tau$ as normalization mode. A systematic uncertainty of 0.35\% per track, for the charged particles having transverse momentum $P_t>200$ MeV/$c$, is determined from a study of partially reconstructed $D^{*+}\rightarrow D^0\pi^+$, $D^0\rightarrow K_S^0 \pi^+\pi^-$, and $K_S^0\rightarrow \pi^+\pi^-$ decays. The systematic uncertainty due to the lepton identification is estimated from the study of $\gamma\gamma\to l^+l^-$ production. An additional uncertainty due to the hadronic interaction is accounted for using the inclusive $J/\psi \to l^+l^-$ events from $B$ decays. We assign a systematic uncertainty of 2.25\% per lepton due to the lepton identification. An uncertainty of 0.32\% is assigned, accounting for the limited MC statistics used for the signal reconstruction efficiency $(\epsilon)$ evaluation. The uncertainty on the number of $B_s^0\bar{B}^0_s$ mesons is $^{+10.3}_{-10.7}\%$, which inherently accounts for the uncertainties on total integrated luminosity, $\sigma_{b\bar{b}}^{\Upsilon(5S)}$, and $f_s$. The uncertainties on the secondary branching fractions are also considered \cite{pdglive}. The systematic uncertainty due to the $J/\psi$ mass-vertex $\chi^2$ requirement, as obtained from the control sample, is 2.25\%.
\begin{table}[h!]
\renewcommand{\arraystretch}{1.5}
    \centering
    \begin{tabular}{p{4cm}|>{\centering\arraybackslash}p{3cm}}
        \hline
        Source & Uncertainty (Events) \\
        \hline\hline
        PDF Parametrization & $^{+0.7}_{-0.8}$\\
        Fit Bias & $^{+0.2}_{-0.0}$ \\
        \hline\hline
        Total (quadratic sum) & $^{+0.7}_{-0.8}$\\
        \hline
    \end{tabular}
    \caption{Additive systematic uncertainties on $\mathcal{B}(B_s^0\rightarrow J/\psi\pi^0$).}
    \label{table:additive}
\end{table}
\begin{table}[h!]
    \renewcommand{\arraystretch}{1.5}
    \centering
    \begin{tabular}{p{4.5cm}|>{\centering\arraybackslash}p{2.5cm}}
        \hline
	    Source & Uncertainty(\%) \\
		\hline\hline
		$\pi^0$ reconstruction & 2.2 \\
		Tracking	& $2\times 0.35$ \\
		Lepton-ID selection & $2\times 2.25$ \\
		MC statistics & 0.32 \\
		Number of $B_s^0$ mesons & $^{+10.3}_{-10.7}$ \\
		$\mathcal{B}(J/\psi\to l^+l^-)$ & 0.77\\
		$\mathcal{B}(\pi^0\to \gamma\gamma)$ & 0.03\\
		$J/\psi$ mass-vertex fit $\chi^2<60$ & 2.25\\
		\hline\hline
		Total (quadratic sum) & $^{+11.7}_{-12.1}$\\
		\hline	
    \end{tabular}
    \caption{Multiplicative systematic uncertainties on $\mathcal{B}(B_s^0\rightarrow J/\psi\pi^0$).}
    \label{table:multiplicative}
\end{table}
\section{VI. UPPER LIMIT ESTIMATION}
With the absence of any significant signal yield, an upper limit (UL) on the branching fraction is calculated using the Bayesian approach. We integrate the profile likelihood ratio $\left(\mathcal{L}_s/\mathcal{L}_\text{max}\right)$ over a range corresponding to physical values of the signal yield, where $\mathcal{L}_s$ is the profile likelihood for a hypothesis of signal yield $s$ and $\mathcal{L}_\text{max}$ is the maximum likelihood of data-fit. The integration is performed from 0\% to 90\% of the total area under the likelihood curve to calculate the UL on the branching fraction. The profile likelihood ratio is convolved with a Gaussian function with a mean of zero and width equal to 0.8 events to incorporate the additive uncertainty in Table \ref{table:additive}. In order to include the systematic uncertainties in the denominator of Eq.~\ref{equ:upperlimitBR} for branching fraction, the modified likelihood ratio is further convolved with a width proportional to the signals, where the total multiplicative systematic uncertainty shown in Table \ref{table:multiplicative} is the proportionality constant. ULs on the yields at 90\% confidence level (CL) are estimated to be 8.03 and 7.64 $B_s^0\to J/\psi\pi^0$ events with and without the systematic uncertainty, respectively. The upper limit on the branching fraction is calculated as
\begin{equation}
\mathcal{B}(B_s^0\to J/\psi\pi^0) = \frac{N_\text{sig}^\text{Yield}\text{ (at 90\% CL)}}{2\times N_{B_s^0\bar{B}_s^0}\times\epsilon\times\mathcal{B}_{J/\psi}\times\mathcal{B}_{\pi^0}},
\label{equ:upperlimitBR}
\end{equation}
where $N_\text{sig}^\text{Yield}$ is the signal yield at 90\% CL, $N_{B^0_s\bar{B}^0_s} = (9.08^{+0.94}_{-0.98})\times 10^6$ is the number of $B^0_s\bar{B}^0_s$ pairs at the $\Upsilon(5S)$ resonance, $\epsilon = 0.310\pm 0.001$ is the signal reconstruction efficiency determined using the MC simulation, $\mathcal{B}_{J/\psi}$ is the sum of $\mathcal{B}(J/\psi\to \mu^+\mu^-)$ and $\mathcal{B}(J/\psi\to e^+e^-)$ \cite{pdglive}, and $\mathcal{B}_{\pi^0}$ is the branching fraction of $\pi^0\to\gamma\gamma$ \cite{pdglive}. The resulting UL with the systematics on $\mathcal{B}(B_s^0\to J/\psi\pi^0)$ at 90\% CL is $1.21 \times 10^{-5}$.

\section{VII. SUMMARY}
In summary, we analyzed the 121.4 fb$^{-1}$ of $e^-e^+$ collision data at the $\Upsilon(5S)$ resonance to search for the decay $B_s^0\to J/\psi\pi^0$. As no signals are observed, we set a UL on $\mathcal{B}(B_s^0\to J/\psi\pi^0)$ of $1.21\times10^{-5}$ at 90\% CL. The reported UL is the most stringent limit and improves the previous upper bound by two orders of magnitude \cite{l3report}. 

\section*{ACKNOWLEDGEMENTS}
This work, based on data collected using the Belle detector, which was
operated until June 2010, was supported by 
the Ministry of Education, Culture, Sports, Science, and
Technology (MEXT) of Japan, the Japan Society for the 
Promotion of Science (JSPS), and the Tau-Lepton Physics 
Research Center of Nagoya University; 
the Australian Research Council including grants
DP210101900, 
DP210102831, 
DE220100462, 
LE210100098, 
LE230100085; 
Austrian Federal Ministry of Education, Science and Research (FWF) and
FWF Austrian Science Fund No.~P~31361-N36;
National Key R\&D Program of China under Contract No.~2022YFA1601903,
National Natural Science Foundation of China and research grants
No.~11575017,
No.~11761141009, 
No.~11705209, 
No.~11975076, 
No.~12135005, 
No.~12150004, 
No.~12161141008, 
and
No.~12175041, 
and Shandong Provincial Natural Science Foundation Project ZR2022JQ02;
the Czech Science Foundation Grant No. 22-18469S;
Horizon 2020 ERC Advanced Grant No.~884719 and ERC Starting Grant No.~947006 ``InterLeptons'' (European Union);
the Carl Zeiss Foundation, the Deutsche Forschungsgemeinschaft, the
Excellence Cluster Universe, and the VolkswagenStiftung;
the Department of Atomic Energy (Project Identification No. RTI 4002), the Department of Science and Technology of India,
and the UPES (India) SEED finding programs Nos. UPES/R\&D-SEED-INFRA/17052023/01 and UPES/R\&D-SOE/20062022/06; 
the Istituto Nazionale di Fisica Nucleare of Italy; 
National Research Foundation (NRF) of Korea Grant
Nos.~2016R1\-D1A1B\-02012900, 2018R1\-A2B\-3003643,
2018R1\-A6A1A\-06024970, RS\-2022\-00197659,
2019R1\-I1A3A\-01058933, 2021R1\-A6A1A\-03043957,
2021R1\-F1A\-1060423, 2021R1\-F1A\-1064008, 2022R1\-A2C\-1003993;
Radiation Science Research Institute, Foreign Large-size Research Facility Application Supporting project, the Global Science Experimental Data Hub Center of the Korea Institute of Science and Technology Information and KREONET/GLORIAD;
the Polish Ministry of Science and Higher Education and 
the National Science Center;
the Ministry of Science and Higher Education of the Russian Federation, Agreement 14.W03.31.0026, 
and the HSE University Basic Research Program, Moscow; 
University of Tabuk research grants
S-1440-0321, S-0256-1438, and S-0280-1439 (Saudi Arabia);
the Slovenian Research Agency Grant Nos. J1-9124 and P1-0135;
Ikerbasque, Basque Foundation for Science, and the State Agency for Research
of the Spanish Ministry of Science and Innovation through Grant No. PID2022-136510NB-C33 (Spain);
the Swiss National Science Foundation; 
the Ministry of Education and the National Science and Technology Council of Taiwan;
and the United States Department of Energy and the National Science Foundation.
These acknowledgements are not to be interpreted as an endorsement of any
statement made by any of our institutes, funding agencies, governments, or
their representatives.
We thank the KEKB group for the excellent operation of the
accelerator; the KEK cryogenics group for the efficient
operation of the solenoid; and the KEK computer group and the Pacific Northwest National
Laboratory (PNNL) Environmental Molecular Sciences Laboratory (EMSL)
computing group for strong computing support; and the National
Institute of Informatics, and Science Information NETwork 6 (SINET6) for
valuable network support.

\end{document}